\documentclass[amsmath, amssymb, amsfonts, twocolumn, footinbib]{revtex4-1}
\usepackage[german,american,english]{babel}
\usepackage{graphicx}
\usepackage{graphics}
\usepackage{dcolumn}
\usepackage{bm}
\usepackage{amssymb}
\usepackage{amsmath}
\usepackage{amsfonts}
\usepackage{epsfig}
\usepackage{mathbbol}
\usepackage{latexsym}
\usepackage{dcolumn}
\usepackage{subfigure}
\usepackage{hyperref}
\usepackage{float}
\usepackage[normalem]{ulem}
\usepackage[usenames, dvipsnames]{color}
\usepackage{epstopdf}

\hypersetup{                    
    colorlinks,
    citecolor=blue,
    filecolor=blue,
    linkcolor=blue,
    urlcolor=blue
}

\newcommand {\pd} [2] {\frac{\partial #1}{\partial #2}}

 \newcommand {\beq}{\begin{equation}}
\newcommand {\eeq}{\end{equation}}
\newcommand {\beqn}{\begin{eqnarray}}
\newcommand {\eeqn}{\end{eqnarray}}
\newcommand {\bit}{\begin{itemize}}
\newcommand {\eit}{\end{itemize}}

\newcommand{\ba}{\begin{array}{rl}}
\newcommand{\ea}{\end{array}}
\newcommand{\bc}{\begin{cases}}
\newcommand{\ec}{\end{cases}}

\newcommand{\om}{\iffalse}

\definecolor{mygray}{gray}{0.6}
\definecolor{gold}{RGB}{150, 150, 10}
\definecolor{mygreen}{RGB}{40, 200, 100}

\newcommand{\bw}[1]{{\textcolor{blue}{#1}}}

\begin{document}
\title{Strong influence of spin-orbit coupling on magnetotransport in two-dimensional hole systems}

\author{Hong Liu, E. Marcellina, A.~R.~Hamilton and Dimitrie Culcer}
\affiliation{School of Physics and Australian Research Council Centre of Excellence in Low-Energy Electronics Technologies, UNSW Node, The University of New South Wales, Sydney 2052, Australia}
\begin{abstract}
With a view to electrical spin manipulation and quantum computing applications, recent significant attention has been devoted to semiconductor hole systems, which have very strong spin-orbit interactions. However, experimentally measuring, identifying, and quantifying spin-orbit coupling effects in transport, such as electrically-induced spin polarizations and spin-Hall currents, are challenging. Here we show that the magnetotransport properties of two-dimensional (2D) hole systems display strong signatures of the spin-orbit interaction. Specifically, the low-magnetic field Hall coefficient and longitudinal conductivity contain a contribution that is second order in the spin-orbit interaction coefficient and is non-linear in the carrier number density. We propose an appropriate experimental setup to probe these spin-orbit dependent magnetotransport properties, which will permit one to extract the spin-orbit coefficient directly from the magnetotransport.
\end{abstract}
\date{\today}
\maketitle

Low-dimensional hole systems have attracted considerable recent attention in the context of nanoelectronics and quantum information \cite{Cuan-2015-EPL, Biswas-2015-EP, Zwanenburg-2013-RMP, Conesa-Boj-2017-NL, Matthias-Zwanenburg-2016-APL, Brauns-Zwanenburg-2016-PRB,Mueller-Zwanenburg-2015-NL, Fanming-Kouwenhoven-2016-NL,Jo-2017-PRB}. They exhibit strong spin-orbit coupling but a weak hyperfine interaction, which allows fast, low-power electrical spin manipulation \cite{Bulaev-2005-PRL,Nichele-Kouwenhoven-2017-PRL} and potentially increased coherence times \cite{Korn-2010-NJP, Salfi-2016-PRL, Salfi-2016-Nanotechnology,Petta-Charlie-Marcus-2005-Sci} while their effective spin-3/2 is responsible for physics inaccessible in electron systems \cite{Roland, Moriya-2014-PRL, Tutul-2014-AP, Shanavas-2016-PRB, Akhgar-2016-NL}. Structures with strong spin-orbit interactions coupled to superconductors may enable topological superconductivity hosting Majorana bound states and non-Abelian particle statistics relevant for topological quantum computation \cite{Lutchyn-2010-PRL, Alicea-2011-NP, Gill-2016-APL,Alestin-India-arXiv}. In the past fabricating high-quality hole structures was challenging, but recent years have witnessed extraordinary experimental progress \cite{Manfra-2005-APL, Habib2009, Hao-2010-NL, Chesi-2011-PRL, Daisy-2016-NL, Srinivasan-2017-PRL,Korn-2010-NJP, Watson-2011-PRB, Srinivasan-2016-RPB,Papadakis-Shayegan-2000-PRL,Shayegan-2002-PRB,Joost-2014-NL,Nichele-Ensslin-2014-PRB,Ota-Tarucha-2004-PRL,Ono-Tarucha-2002-Sci,Tarucha-2007-RMP,Croxall2013,Katsaros-2016-NL,Gerl2005,Clarke2007}. 

A full quantitative understanding of spin-orbit coupling mechanisms is vital for the realization of spintronics devices and quantum computation architectures \cite{Das-2004-RMP, Awschalom-2007-NP}. At the same time experimental measurement of spin-orbit parameters is difficult \cite{SasakiA-2014-NN}. Spin-orbit constants can be estimated from weak antilocalization \cite{Kallaher-2010-PRB,Boris-2008-PRB,Nakamura-2012-PRL,Koga-2002-PRL}, Shubnikov-de Haas oscillations and spin precession in large magnetic fields (up to 2 T) \cite{Nitta-1997-PRL,Youn-2013-APL,Li2016}, and state-of-the-art optical measurements \cite{Eldridge-2010-AIP,Wang-2013-NC}. Many techniques yield only the ratio between the Rashba and Dresselhaus terms or allow the determination of only one type of spin splitting. Likewise, experimentally quantifying spin-orbit induced effects, such as via spin-to-charge conversion or vice versa, is difficult. For instance, current-induced spin polarizations in spin-orbit coupled systems are small and their relationship to theoretical estimates is ambiguous \cite{Chao-Xing-2008-PRB,Tokatly-2015-PRB,Li-2016-NC}, while spin-Hall currents \cite{Wong-2010-PRB} can only be identified via an edge spin accumulation \cite{Sonin-2010-PRB,Nomura-2005-PRB,Jungwirth-2012-NM}.

\begin{figure}
\begin{center}
\includegraphics[width=0.9\columnwidth]{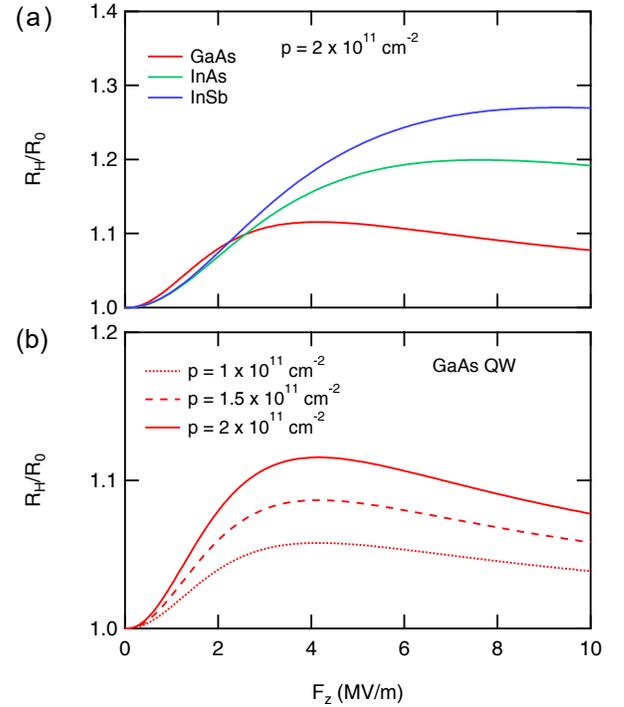}
\caption{\label{RH}Spin-orbit correction to the Hall coefficient $R_H$ of 2D holes in various 15 nm quantum wells as a function of the electric field $F_z$ across the well, where $R_0\equiv\frac{1}{pe}$ is the bare Hall coefficient. Panel shows results for (a) different quantum well materials at $p = 1 \times 10^{11}$ cm$^{-2}$ and (b) GaAs quantum wells at different densities.}
\end{center}
\end{figure}

Here we show that the spin-orbit interaction can have a sizeable effect on low magnetic-field Hall transport in a 2D hole system, which is density-dependent and experimentally visible. Our central result, shown in Fig.~\ref{RH}, is a correction to the low-field Hall coefficient
\begin{equation}\label{RH}
R_H =\frac{1}{pe}\bigg[1 + \bigg(\frac{64\pi m^{*2}\alpha^2}{\hbar^4}\bigg) \, p\bigg],
\end{equation}
where $\alpha$ is the coefficient of the cubic Rashba spin-orbit term, which arises from the application of an electric field $F_z$ across the quantum well, $m^*$ is the heavy-hole effective mass at $\alpha = 0$, $p$ is the hole density, and $e$ is the elementary charge. Note that here we have chosen the $z-$axis as the quantization direction. In hole systems, where the spin-orbit coupling can account for as much as 40$\%$ of the Fermi energy \cite{Elizabeth-2017-PRB}, effects of second-order in the spin-orbit strength can be sizable in charge transport. These reflect spin-orbit corrections to the occupation probabilities, density of states, and scattering probabilities, as well as the feedback of the current-induced spin polarization on the charge current. Quantitative evaluation shows that the spin-orbit corrections can reach more than $10\%$ in GaAs quantum wells, and are of the order $\sim20-30\%$ in InAs and InSb quantum wells (Fig.~\ref{RH}a). The magnitude of the spin-orbit corrections also increase with density, which is consistent with the expectation that the strength of spin-orbit interaction increases with density (Fig.~\ref{RH}b). It is worth noting that the correction due to spin-orbit coupling has already taken into account the fact that the spin-split subbands may have different hole mobilities.

In the following we derive the formalism and show how spin-orbit coupling can give rise to corrections in the magnetotransport. We consider a 2D hole system in the presence of a constant electric field ${\bm F}$ and a perpendicular magnetic field ${\bm B} = B_z \hat{\bm z}$. The full Hamiltonian is $\hat{H} = \hat{H}_0 + \hat{H}_E + \hat{U} + \hat{H}_{Z}$, where the band Hamiltonian $\hat{H}_0$ is defined below in Eq.~(\ref{Band-H}), $\hat{H}_E = -e{\bm F}\cdot\hat{\bm r}$ represents the interaction with the external electric field of holes  $\hat{\bm r}$ is the position operator, and $\hat{U}$ is the impurity potential, discussed below. The Zeeman term $H_{Z} = 3\kappa \mu_B {\bm \sigma}\cdot{\bm B}$ with $\kappa$ is a material-specific parameter \cite{Roland}, $\mu_B$ the Bohr magneton and ${\bm \sigma}$ the vector of Pauli spin matrices. Rashba spin-orbit coupling is expected to dominate greatly over the Dresselhaus term in 2D hole gases, even in materials such as InSb in which the bulk Dresselhaus term is very large \cite{Elizabeth-2017-PRB}. With this in mind, the band Hamiltonian used in our analysis in the absence of a magnetic field is written as \cite{Winkler-2000-PRB}
\begin{equation}\label{Band-H} 
H_{0{\bm k}}=\frac{\hbar^2 k^2}{2m^*} + i\alpha(k^3_-\sigma_+ - k^3_+\sigma_+) \equiv \frac{\hbar^2 k^2}{2m^*} + {\bm \sigma} \cdot {\bm \Omega}_{\bm k},
\end{equation} 
where $m^* = \frac{m_0}{\gamma_1+\gamma_2}$, the Pauli matrix $\sigma_{\pm}=\frac{1}{2}(\sigma_x\pm i\sigma_y)$, $k_{\pm} = k_x\pm i k_y$. For $B = 0$ the eigenvalues of the band Hamiltonian are $\varepsilon_{{\bm k}\pm} = \hbar^2 k^2/(2m^*) \pm \alpha k^3$. In an external magnetic field we replace ${\bm k}$ by the gauge-invariant crystal momentum $\tilde{{\bm k}} = {\bm k} - e{\bm A}$ with the vector potential ${\bm A} = \frac{1}{2}(-y, x, 0)$. The magnetic field is assumed small enough that Landau quantization can be neglected, in other words $\omega_c {\tau_p} \ll 1$, where $\omega_c = eB_z/m^*$ is the cyclotron frequency and $\tau_p$ the momentum relaxation time. 

To set up our transport formalism, in the spirit of Ref.~[\onlinecite{Vasko}], we begin with a set of time-independent states $\{{\bm k}s\}$, where $s$ represents the twofold heavy-hole pseudospin. We work in terms of the canonical momentum $\hbar {\bm k}$. The terms $\hat{H}_0$, $\hat{H}_{E}$ and $\hat{H}_{Z}$ are diagonal in wave vector but off-diagonal in band index while for elastic scattering in the first Born approximation $U^{ss'}_{{\bm k}{\bm k}'} =U_{{\bm k}{\bm k}'}\delta_{ss'}$. Without loss of generality, here we consider short-range impurity scattering. The impurities are assumed uncorrelated and the average of $\langle {\bm k}s|\hat{U}|{\bm k}'s'\rangle\langle{\bm k}'s'\hat{U}|{\bm k}s\rangle$ over impurity configurations is  $(n_i |\bar{U}_{{\bm  k}'{\bm k}}|^2\delta_{ss'})/V$, where $n_i$ is the impurity density, $V$ the crystal volume, and $\bar{U}_{{\bm  k}'{\bm k}}$ the matrix element of the potential of a single impurity. 

The central quantity in our theory is the density operator $\hat{\rho}$, which satisfies the quantum Liouville equation,
\begin{equation} 
\frac{d \hat{\rho}}{dt}+\frac{i}{\hbar}[\hat{H},\hat{\rho}]=0.
 \end{equation}
The matrix elements of  $\hat{\rho}$ are $\hat{\rho}_{{\bm k}{\bm k}'}\equiv \hat{\rho}^{ss'}_{{\bm k}{\bm k}'}=\langle {\bm k}s |\hat{\rho}|{\bm k}'s'\rangle$ with understanding 
that $\hat{\rho}_{{\bm k}{\bm k}'}$  is a matrix in heavy hole subspace. The density matrix  $\rho_{{\bm k}{\bm k}'}$  is written as  $\rho_{{\bm k}{\bm k}'}=f_{\bm k}\delta_{{\bm k}{\bm k}'}+g_{{\bm k}{\bm k}'}$, where $f_{\bm k}$ is diagonal in wave vector, while $g_{{\bm k}{\bm k}'}$ is off-diagonal in wave vector. The quantity of interest in determining the charge current is $f_{\bm k}$ since the current operator is diagonal in wave vector. We therefore derive an effective equation for this quantity by first breaking down the quantum Liouville equation into the kinetic equations of $f_{\bm k}$ and $g_{{\bm k}{\bm k}'}$ separately, and $f_{\bm k}$ obeys
  \begin{equation}\label{Kinetic-e} 
  \frac{df_{\bm k}}{dt}+\frac{i}{\hbar}\left[H_{0{\bm k}} + H_Z, f_{\bm k}\right]+\hat{J}(f_{\bm k})=\mathcal{D}_{E,{\bm k}}+\mathcal{D}_{L,{\bm k}},
  \end{equation}
where the scattering term in the Born approximation
  \begin{equation} \label{J0}
  \hat{J}(f_{\bm k})\!=\!\frac{1}{\hbar^2}\int^{\infty}_0\!dt'[\hat{U},e^{-\frac{iH_0t'}{\hbar}}[\hat{U},\hat{f}(t)]e^{\frac{iH_0t'}{\hbar}}]_{{\bm k}{\bm k}},
  \end{equation}  
and the driving terms
\begin{subequations}\label{Driving-term}
\begin{equation}\label{DE}
\mathcal{D}_{E,{\bm k}} = -\frac{e{\bm E}}{\hbar}\cdot\pd{f_{{\bm k}}}{{\bm k}},
\end{equation}
\begin{equation}\label{DL}
\mathcal{D}_{L,{\bm k}} = \frac{1}{2}\frac{e}{\hbar}\{\hat{\bm v}\times {\bm B}, \pd{f_{{\bm k}}}{{\bm k}}\},
\end{equation}
\end{subequations}
stem from the applied electric field and Lorentz force respectively ~[\onlinecite{Vasko}]. In external electric and magnetic fields one may decompose $f_{\bm k} = f_{0{\bm k}} + f_{E{\bm k}} + f_{EB{\bm k}}$, where $f_{0{\bm k}}$ is the equilibrium density matrix, $f_{E{\bm k}}$ is a correction to first order in the electric field (but at zero magnetic field), and $f_{EB{\bm k}}$ is an additional correction \bw{that is} first order in the electric and magnetic fields. The equilibrium density matrix is written as $f_{0{\bm k}} = (1/2) \, [(f_{{\bm k}+} + f_{{\bm k}-}) \openone + {\bm \sigma}\cdot\hat{\bm \Omega} (f_{{\bm k}+} - f_{{\bm k}-})]$, where $\hat{\bm \Omega}$ is a unit vector and ${\bm \Omega}$ was defined in Eq.~(\ref{Band-H}), and $f_{{\bm k}\pm}$ represent the Fermi-Dirac distribution functions corresponding to the two band energies $\varepsilon_{{\bm k}\pm}$. In linear response one may replace $f_{{\bm k}} \rightarrow f_{0{\bm k}}$ in Eq.~(\ref{DE}). On the other hand it is trivial to check that the driving term $\mathcal{D}_{L,{\bm k}}$ vanishes when the equilibrium density matrix is substituted, so in Eq.~(\ref{DL}) one may replace $f_{\bm k} \rightarrow f_{E{\bm k}}$. Hence, in this work we perform a perturbation expansion up to first order in the electric and magnetic fields, and up to second order in the spin-orbit interaction, retaining terms up to order $\alpha^2$. The detailed solution of Eq.~(\ref{Kinetic-e}) and the explicit evaluation of the scattering term Eq.~(\ref{J0}) are given in the Supplement. We briefly summarize the procedure here. Firstly, with $f_{0{\bm k}}$ known and only $\mathcal{D}_{E,{\bm k}}$ on the right-hand side of Eq.~(\ref{Kinetic-e}), we obtain $f_{E{\bm k}}$. Next, with only $\mathcal{D}_{L, {\bm k}}$ on the right-hand side of Eq.~(\ref{Kinetic-e}), we obtain $f_{EB{\bm k}}$. By taking the trace with current operator the longitudinal and transverse components of the current are found as $ j_{x, y}=e\text{Tr}\big[\hat{v}_{x, y} f_{{\bm k}}\big]$, with $v_i = (1/\hbar) \, \partial H_{0{\bm k}}/\partial {\bm k}$. Finally, with $\sigma_{xx}$ and $\sigma_{xy}$ the longitudinal and Hall conductivities respectively, the Hall coefficient appearing in Eq.~(\ref{RH}) is found through $R_H=\frac{\sigma_{xy}}{B_z(\sigma^2_{xx} + \sigma^2_{xy})}$. For the Hall conductivity on the other hand one needs $f_{EB{\bm k}}$. We note that the topological Berry curvature terms that give contributions analogous to the \textit{anomalous} Hall effect in Rashba systems (with the magnetization replaced by the magnetic field $B_z$) vanish identically when both the band structure and the disorder terms are taken into account. 

\begin{table}[H]
	\centering
	\caption{\label{Density-window}The maximal hole densities for which the current theory is applicable for 15 nm-wide GaAs, InAs, and InSb quantum wells. Densities in units of $10^{11}\text{cm}^{-2}$.}
	\begin{tabular}{|m{2.5cm}|m {2.5cm}|m{2.5cm}|}
		\hline
		GaAs & InAs & InSb \\
		\hline
		$6.55$  &  $8.08$ & $8.60$ \\
		\hline
	\end{tabular}
\end{table}

The limits of applicability of our approach are as follows. We assume that the magnetotransport considered here occurs in the weak disorder regime, i.e. $\varepsilon_{\text{F}}\tau_p/\hbar \gg 1 $, where $\varepsilon_{\text{F}}$ is Fermi energy. Furthermore, we assume that the scattering does not change appreciably when the gate field is changed at low density \cite{Croxall2013}, so the condition $\varepsilon_{\text{F}}{\tau_p}/\hbar \gg 1$ is still valid when the gate field is changed. We assume $\alpha k^3_{\text{F}}/\epsilon_{kin} \ll 1$ where $\epsilon_{kin}=\frac{\hbar^2k^2_{\text{F}}}{2m^*}$ is kinetic energy, for example in Ref.~[\onlinecite{Boris-2008-PRB}], the spin-orbit-induced splitting of the heavy hole sub-band at the Fermi level is determined to be around $30\%$ of the total Fermi energy. In addition, Eq.~(\ref{Band-H}) with $\alpha$ independent of wave vector is a result of the Schrieffer-Wolff transformation applied to the Luttinger Hamiltonian, and its use requires the Schrieffer-Wolff method to be applicable. Furthermore, throughout this paper we consider cases where only the HH1 band is occupied. We have calculated the exact window of applicability of our theory in Table~\ref{Density-window}.

Physically, the terms $\propto \alpha^2$ entering the Hall coefficient are traced back to several mechanisms. Firstly, spin-orbit coupling gives rise to corrections to: (i) the occupation probabilities, through $f_{{\bm k}\pm}$; (ii) the band energies and density of states, through $d\varepsilon_{{\bm k}\pm}/dk$; and (iii) the scattering term, which includes intra- and inter-band scattering, as well as scattering between the charge and spin distributions. Secondly, Rashba spin-orbit coupling gives rise to a current-induced spin polarization \cite{Chao-Xing-2008-PRB}, which is of first order in $\alpha$, and this in turn gives rise to a feedback effect on the charge current, which is \bw{then} responsible for approximately a quarter of the overall spin-orbit contribution to the Hall coefficient. 

As a concrete example, a 2D hole system confined to GaAs/AlGaAs heterostructures is particularly promising since it has not only a very high mobility, but also a spin splitting that has been shown to be electrically tunable in both square and triangular wells \cite{Lu-1998-PRL}. The spin splitting can be tuned from large values to nearly zero in a square quantum well whose charge distribution can be controlled from being asymmetric to symmetric via the application of a surface-gate bias. Whereas thus far the theoretical formalism has been general, to make concrete experimental predictions we first specialize to a two-dimensional hole gas (2DHG) in a 15 nm-wide GaAs quantum well subjected to an electric field in the $\hat{z}$ direction, so that the symmetry of the quantum well can be tuned arbitrarily. In the simplest approximation, taking into account only the lowest heavy-hole and light-hole sub-bands, in a 2DHG the Rashba coefficient $\alpha$ may be estimated as
\begin{equation}
\alpha = \frac{3\hbar^4}{m^2_0\Delta_E} \overline{\gamma}^2\langle \phi_L|\phi_H\rangle\langle\phi_H| (- i \, d/dz) |\phi_L\rangle.
\end{equation} 
where $\Delta_E$ is energy splitting of the lowest heavy-hole and light-hole sub-bands and $\overline{\gamma}=\frac{\gamma_2+\gamma_3}{2}$, and $\phi_{H,L} \equiv \phi_{H,L} (z)$ represents the \textit{orbital} component of the heavy-hole and light-hole wave functions respectively in the direction $\hat{\bm z}$ perpendicular to the interface. For a system with top and back gates, where the electric field $F_z$ across the well can be turned on or off, we use a modified infinite square well wave function in which $F_z$ is already encoded \cite{Bastard-1983-PRB}.

\begin{figure}
	\begin{center}
		\includegraphics[width=1.0\columnwidth,scale=0.9]{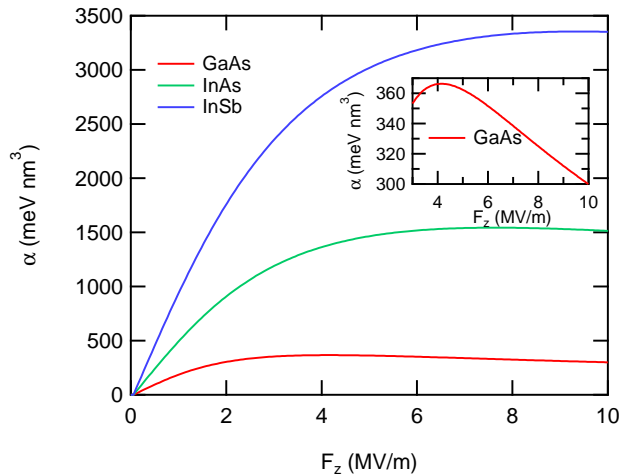}
		\caption{\label{Rashba}The Rashba coefficient $\alpha$ of as a function of the net perpendicular electric field $F_z$ for 15 nm GaAs, InAs, and InSb quantum wells. The inset shows that $\alpha$ for GaAs decreases by $\sim 20\%$ as $F_z$ is increased from $4$ MV/m to $10$ MV/m, due to the fact the well becomes quasi-triangular at $F_z \gtrsim 4$ MV/m.}
	\end{center}
\end{figure}

The Rashba coefficient $\alpha$, as a function of $F_z$, for 15 nm hole quantum wells is shown Fig. \ref{Rashba}. For GaAs, at low $F_z$ ($F_z \ll 4$ MV/m), the Rashba coefficient increases with $F$, which is in accordance with the trends reported in Ref. \cite{Papadakis-1999-Science}. As $F_z$ is increased, $\alpha$ then saturates, and, at larger electric fields ($F_z > 4$ MV/m), the quantum well becomes quasi-triangular and the Rashba coefficient $\alpha$ decreases with increasing electric field $F_z$. The decrease of $\alpha$ as a function of $F_z$ in quasi-triangular wells is consistent with the experimental findings of Ref.~[\onlinecite{Habib-2004-APL}]. Note that for different materials, $\alpha$ saturates at different values of $F_z$, and that the $\alpha$ is larger in materials with a higher atomic number \cite{Elizabeth-2017-PRB}.
	
Given the dependence of $\alpha$ (Fig. \ref{Rashba}), and hence the Hall coefficient $R_H$ (Fig. \ref{RH}), on $F_z$, we now outline how $\alpha$ can be deduced experimentally. Using a top- and backgated quantum well, the quantum well is initially tuned to be symmetric so that $\alpha$ will be zero and the hole density can be measured accurately. One subsequently increases $F_z$, for example to $\sim$ 4 MV/m for the GaAs quantum well discussed above, whilst keeping the density constant. This in turn results in an appreciable increase in $\alpha$, and hence a large change in $R_H$ as a function of $F_z$.

\begin{figure}
	\begin{center}
		\includegraphics[width=0.9\columnwidth]{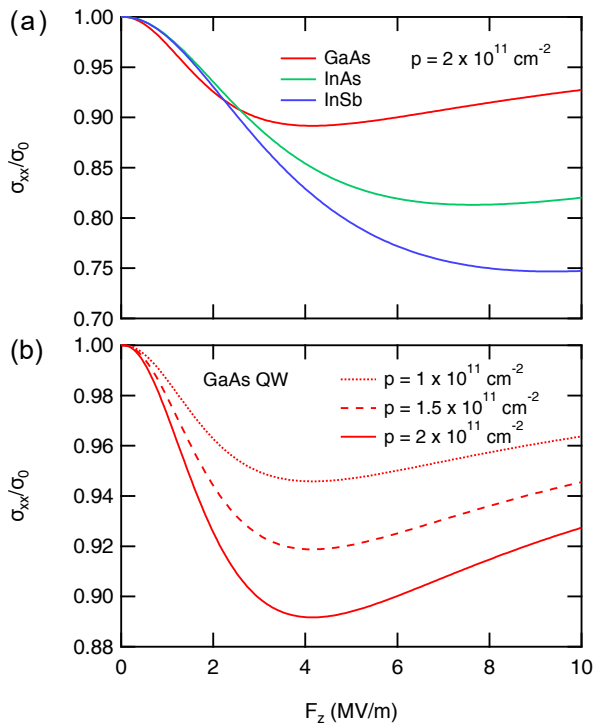}
		\caption{\label{longit}Ratio of Drude conductivity at finite electric fields to its zero electric field value, with the bare Drude conductivity $\sigma_0\equiv pe\mu$, for (a) different quantum well materials at $p = 1 \times 10^{11}$ cm$^{-2}$ and (b) GaAs quantum wells at different densities. Here, the well width is 15 nm.}
	\end{center}
\end{figure}

The non-monotonic change in $\alpha$ as a function of $F_z$ likewise affects the longitudinal conductivity $\sigma_{xx}$  (Fig.~\ref{longit}), which reads
\begin{equation}\label{sigma_xx}
\sigma_{xx} =\sigma_0\bigg[1 - \bigg(\frac{60\pi m^{*2}\alpha^2}{\hbar^4}\bigg) \, p\bigg].
\end{equation}
The spin-orbit corrections are larger in InAs and InSb (Fig.~\ref{longit}a) rather than GaAs. Furthermore, as the density increases, $\sigma_{xx}$ decreases faster with $F_z$ (Fig.~\ref{longit}b). However, although the spin-orbit corrections to $\sigma_{xx}$ have a similar functional form as and a similar magnitude to the corrections to $R_H$, it is difficult to single out the dependence of $\sigma_{xx}$ on $\alpha$ experimentally. As the shape of the wave functions changes with $F_z$, the spin-orbit independent scattering properties are also altered, which may then introduce a larger correction to $\sigma_{xx}$ than the spin-orbit induced corrections \cite{Simmons1997}. In fact, the spin-orbit independent corrections can alter the carrier mobility by $\sim 20\%$ even in electron quantum wells \cite{Croxall2013}.

Various possibilities exist to extend the scope of the calculations presented in this paper. Here we have restricted ourselves, for the sake of gaining physical insight and without loss of generality, to hole systems in which the Schrieffer-Wolff approximation is applicable so that $\alpha$ can be approximated as constant. In a general 2D hole system $\alpha(k)$ is a function of wave vector, and decreases with $k$ at larger wave vectors. Its behaviour is in principle not tractable analytically though it can straightforwardly be calculated numerically. The results we have found remain true in their general closed form for hole systems at arbitrary densities provided $\alpha$ is replaced by $\alpha(k)$. An alternative approach would be to start directly with the $4\times4$ Luttinger Hamiltonian and determine the charge conductivity using a spin-3/2 model. However, calculating the conductivity as a function of $F_z$ can quickly become very complicated analytically, limiting the utility of such an approach. Finally, the kinetic equation approach we have discussed can straightforwardly be generalized to arbitrary band structures in a way that makes it suitable for fully numerical approaches relying on maximally localized Wannier functions \cite{Culcer_InterbandCoh_PRB2017}.

It is worth mentioning how the corrections in the magnetotransport properties of 2D electrons will differ from those of 2D holes. In 2D electrons, to lowest order the spin-orbit coupling stems from $\boldsymbol{k.p}$ coupling with the topmost valence band, and the leading contribution to spin-orbit interaction in 2D electrons is linear in $k$ \cite{Roland}. As a result, the spin-orbit dependent corrections to the magnetotransport in 2D electrons will be much smaller compared to 2D holes, and thus may not be detectable within experimental resolution.

In summary, we have presented a quantum kinetic theory of magneto-transport in 2D heavy-hole systems in a weak perpendicular magnetic field and demonstrated that the Hall coefficient, as well as the longitudinal conductivity, display strong signatures of the spin-orbit interaction.  We have also shown that our theory provides an excellent qualitative agreement to existing experimental trends for $\alpha$, although to the best of our knowledge, there has not been a demonstration of $R_{H}$ changing as a function of $\alpha$. An appropriate experimental setup with top and back gates can lead to a direct electrical measurement of the Rashba spin-orbit constant via the Hall coefficient.

\acknowledgments
This research was supported by the Australian Research Council Centre of Excellence in Future Low-Energy Electronics Technologies (project CE170100039) and funded by the Australian Government.

\end{document}


\title{Supplement of ``Strong influence of spin-orbit coupling on magnetotransport in two-dimensional hole systems"}
\author{Hong Liu, E. Marcellina, A.~R.~Hamilton and Dimitrie Culcer}

\affiliation{School of Physics and Australian Research Council Centre of Excellence in Low-Energy Electronics Technologies, UNSW Node, The University of New South Wales, Sydney 2052, Australia}
\maketitle
\section{Luttinger Hamiltonian}
We start from the bulk  $4\times4$ Luttinger Hamiltonian \cite{Luttinger-1956-PR} $H_L(k^2,k_z)$ describing holes in the uppermost valence band with an effective spin $J=3/2$. So the hole system with top and back gate in $z$-direction can be simplified as the isotropic Luttinger-Kohn Hamiltonian plus a confining asymmetrical triangular potential.
\begin{equation}
\hat{H}=H_L(k^2,k_z)- e F_z z, \quad  z>0,
\end{equation}
where $F_{z}$ is the gate electric field and $F_z \geq 0$. The $4\times4$ Luttinger Hamiltonian, which is expressed in the basis of $J_z$ eigenstates
$\{|+\frac{3}{2}\rangle,|-\frac{3}{2}\rangle,|+\frac{1}{2}\rangle,|-\frac{1}{2}\rangle\}$, reads
\begin{equation}
\ba
H_L(k^2,k_z)\!=\!
\left(
\begin{array}{cccc}
P+Q & 0  & L& M \\
0  & P+Q  & M^* & -L^*  \\  
L^* & M & P-Q &0 \\
M^* & -L & 0 & P-Q
\end{array}
\right),
\ea
\end{equation}
where 
\begin{equation}
\ba
&\dps P=\frac{\mu}{2}\gamma_1(k^2+k^2_z),\quad Q=-\frac{\mu}{2}\gamma_2(2k^2_z-k^2),\\[3ex]
&\dps L=-\sqrt{3}\mu\gamma_3k_-k_z,\quad  M=-\frac{\sqrt{3}\mu}{2}(\overline{\gamma}k^2_-+\delta k^2_+).
\ea
\end{equation}
with $\mu=\frac{\hbar^2}{m_0}$, $\gamma_1,\gamma_2,\gamma_3$ are the Luttinger parameters (Table~\ref{LP}),
$\overline{\gamma}=\frac{\gamma_2+\gamma_3}{2}$, $\delta=\frac{\gamma_2-\gamma_3}{2}$, and $k^2=\sqrt{k^2_x+k^2_y}$, $k_{\pm}=k_x\pm ik_y$ and $\theta=\arctan \frac{k_y}{k_x}$.
\begin{table}
\centering
\caption{\label{LP}Luttinger parameters used in this work \cite{Roland}.}
\begin{tabular}{|m{2cm}|m {2cm}|m{2cm}|m{2cm}|}
\hline
 & GaAs & InAs & InSb \\
 \hline
$\gamma_1$ & 6.85  &  20.40 & 37.10 \\
\hline
$\gamma_2$ & 2.10  &  8.30 & 16.50 \\
\hline
$\gamma_3$ & 2.90  &  9.10 & 17.70 \\
\hline
\end{tabular}
\end{table}
To obtain the spectrum of our system, we use modified infinite square well wave functions \cite{Bastard-1983-PRB} for the heavy hole (HH) and light hole (LH) states
 \begin{equation}\label{Bastard}
 	\phi_v = \frac{\sin\left[\frac{\pi}{d}\left(z + \frac{d}{2}\right)\right] \exp \left[ -\beta_v \left(\frac{z}{d} + \frac{1}{2} \right) \right]}{\pi \sqrt{\frac{e^{-\beta_v} d \sinh(\beta_v) }{2 \pi^2 \beta_v + 2 \beta_v^3}}}, 
 \end{equation}
where $v = h, l$ denote the HH and LH states and $d$ is the width of the quantum well. The eigenvalues of the heavy hole and light hole as well as the corresponding ${\bm k}$ dependent expansion coefficients are then obtained by diagonalizing the matrix $ \tilde{H}$, whose elements are given as
 \begin{equation}\label{4H}
 \tilde{H}=\langle \nu|H_L(k^2,\hat{k}_z)+V(z)|\nu'\rangle, 
 \end{equation}
where $|\nu\rangle$ denotes the wave function Eq.~(\ref{Bastard}) and $\hat{k}_z$ stands for the operator $-i\pd{}{z}$. 
The two lowest eigenenergies of the $4\times4$ matrix Eq.~(\ref{4H}) correspond to  the dispersion of the spin-split HH1$_{\pm}$ subbands.  Usually, only the lowest HH-subspace is taken into account at low hole densities. Accordingly, we perform a Schrieffer-Wolff transformation on Eq. \ref{4H} to restrict our attention to the lowest HH subspace. Therefore, the effective Hamiltonian describing the two dimensional hole gas is \cite{Winkler-2000-PRB}
\begin{equation}\label{Band-H-App} 
H_{0{\bm k}}=\frac{\hbar^2 k^2}{2m^*}+i\alpha(k^3_-\sigma_+-k^3_+\sigma_+),
\end{equation} 
where $m^*\equiv m^*_{hh}=\frac{m_0}{\gamma_1+\gamma_2}$ and the Pauli matrix $\sigma_{\pm}=\frac{1}{2}(\sigma_x\pm i\sigma_y)$. The eigenvalues of Eq.~(\ref{Band-H-App}) are
$ \varepsilon_{k,\pm}=\epsilon_0\pm \alpha k^3$, where $\epsilon_0=\frac{\hbar^2 k^2}{2m^*}$.
The Rashba coefficient $\alpha$ is expressed as
 \begin{equation}
\alpha = \frac{3\mu^2}{\Delta_E} \overline{\gamma}^2\langle \phi_L|\phi_H\rangle\langle\phi_H|k_z|\phi_L\rangle.
\end{equation} 
where $\Delta_E$ is energy splitting of heavy hole and light hole. 

\section{Scattering term}\label{J}
The ${\bm k}$-diagonal part of density matrix $f_{\bm k}$ is a $2\times2$ Hermitian matrix, which is decomposed into
$f_{\bm k}=n_{\bm k}\openone+S_{\bm k}$, where $n_{\bm k}$ represents the scalar part and $\openone$ is the identity matrix into two dimensions. The component $S_{\bm k}$ is written purely in terms of the Pauli $\sigma$ matrices $S_{\bm k}=\frac{1}{2}{\bm S}_{\bm k}\cdot{\bm \sigma}\equiv \frac{1}{2}S_{{\bm k}i}\sigma_i$. With this notation, the scattering term is in turn decomposed into 
 \begin{equation}
 \ba 
 \hat{J}(f_{\bm k})
&\dps =\frac{n_i}{\hbar^2}\int \frac{d^2{k'}}{(2\pi)^2}|\overline{U}_{{\bm k}{\bm k}'}|^2 (n_{\bm k}-n_{{\bm k}'}) \lim\limits_{\eta \rightarrow 0}
\int^{\infty}_0 dt'e^{-\eta t'}e^{-iH_{0{\bm k}'}t'/\hbar}e^{iH_{0{\bm k}}t'/\hbar}+\text{H.c.}\\[3ex]
&\dps +\frac{n_i}{2\hbar^2}\int \frac{d^2{k'}}{(2\pi)^2}|\overline{U}_{{\bm k}{\bm k}'}|^2 ({\bm S}_{\bm k}-{\bm S}_{{\bm k}'})\cdot \lim\limits_{\eta \rightarrow 0}
\int^{\infty}_0 dt'e^{-\eta t'}e^{-iH_{0{\bm k}'}t'/\hbar}{\bm \sigma}e^{iH_{0{\bm k}}t'/\hbar}+\text{H.c.}.
\ea
\end{equation}
We use perturbation theory solving the kinetic equation up to $\alpha^2$. In the process, we decompose the matrix $S_{\bm k}= S_{{\bm k}\parallel}+S_{{\bm k}\perp}$ and write those two parts as $S_{{\bm k}\parallel}=(1/2)s_{{\bm k}\parallel}\sigma_{{\bm k}\parallel}$ and $S_{{\bm k}\perp}=(1/2)s_{{\bm k}\perp}\sigma_{{\bm k}\perp}$. The terms $s_{{\bm k}\parallel}$ and $s_{{\bm k}\perp}$ are scalars and given by $s_{{\bm k}\parallel}={\bm S}_{\bm k}\cdot \hat{\bm \Omega}_{\bm k}$ and $s_{{\bm k}\perp}={\bm S}_{\bm k}\cdot \hat{\bm \Theta}_{\bm k}$ with
$\hat{\bm \Omega}_{\bm k}=-\sin3\theta \hat{\bm x}+\cos3\theta\hat{\bm y}$ and $\hat{\bm \Theta}_{\bm k}=-\cos3\theta\hat{\bm x}-\sin3\theta\hat{\bm y}$. With $\gamma=\theta'-\theta$, the scattering term becomes
   \begin{equation}
   \ba
  \hat{J}(n) &\dps =\frac{\pi n_i}{2\hbar}\int\frac{d^2k'}{(2\pi)^2}|\overline{U}_{{\bm k}{\bm k}'}|^2(n_{\bm k}-n_{{\bm k}'})
  \cdot(1+\hat{\bm \Omega}_{{\bm k}'}\cdot\hat{\bm \Omega}_{\bm k})\Big[\delta(\epsilon_+-\epsilon'_+)+\delta(\epsilon_--\epsilon'_-)\Big] \\[3ex]
  &\dps +\frac{\pi n_i}{2\hbar}\int\frac{d^2k'}{(2\pi)^2}|\overline{U}_{{\bm k}{\bm k}'}|^2(n_{\bm k}-n_{{\bm k}'})\cdot{\bm \sigma}
  \cdot(\hat{\bm \Omega}_{{\bm k}'}+\hat{\bm \Omega}_{\bm k})\Big[\delta(\epsilon'_+-\epsilon_+)-\delta(\epsilon'_--\epsilon_-)\Big]\\[3ex]
  &\dps +\frac{\pi n_i}{2\hbar}\int\frac{d^2k'}{(2\pi)^2}|\overline{U}_{{\bm k}{\bm k}'}|^2(n_{\bm k}-n_{{\bm k}'})
  \cdot(1-\hat{\bm \Omega}_{{\bm k}'}\cdot\hat{\bm \Omega}_{\bm k})\Big[\delta(\epsilon_+-\epsilon'_-)+\delta(\epsilon_--\epsilon'_+)\Big] \\[3ex]
  &\dps +\frac{\pi n_i}{2\hbar}\int\frac{d^2k'}{(2\pi)^2}|\overline{U}_{{\bm k}{\bm k}'}|^2(n_{\bm k}-n_{{\bm k}'})\cdot{\bm \sigma}
  \cdot \Big[(\hat{\bm \Omega}_{\bm k}-\hat{\bm \Omega}_{{\bm k}'})\Big]\Big[\delta(\epsilon'_--\epsilon_+)-\delta(\epsilon'_+-\epsilon_-)\Big],
      \ea
  \end{equation} 
  and
   \begin{equation}
   \ba 
   \hat{J}(S) &\dps =\frac{\pi n_i}{4\hbar}\int\frac{d^2k'}{(2\pi)^2}|\overline{U}_{{\bm k}{\bm k}'}|^2({\bm S}_{\bm k}-{\bm S}_{{\bm k}'})
  \cdot\Big[{\bm \sigma}(1-\hat{\bm \Omega}_{\bm k}\cdot\hat{\bm \Omega}_{{\bm k}'})+(\hat{\bm \Omega}_{\bm k}\cdot{\bm \sigma})\hat{\bm \Omega}_{{\bm k}'}+\hat{\bm \Omega}_{\bm k}(\hat{\bm \Omega}_{{\bm k}'}\cdot{\bm \sigma})\Big]\Big[\delta(\epsilon_+-\epsilon'_+)+\delta(\epsilon_--\epsilon'_-)\Big] \\[3ex]
  &\dps +\frac{\pi n_i}{4\hbar}\int\frac{d^2k'}{(2\pi)^2}|\overline{U}_{{\bm k}{\bm k}'}|^2({\bm S}_{\bm k}-{\bm S}_{{\bm k}'})
  \cdot(\hat{\bm \Omega}_{\bm k}+\hat{\bm \Omega}_{{\bm k}'})\Big[\delta(\epsilon'_+-\epsilon_+)-\delta(\epsilon'_--\epsilon_-)\Big] \\[3ex]
   &\dps +\frac{\pi n_i}{4\hbar}\int\frac{d^2k'}{(2\pi)^2}|\overline{U}_{{\bm k}{\bm k}'}|^2({\bm S}_{\bm k}-{\bm S}_{{\bm k}'})
  \cdot\Big[{\bm \sigma}(1+\hat{\bm \Omega}_{\bm k}\cdot\hat{\bm \Omega}_{{\bm k}'})-(\hat{\bm \Omega}_{\bm k}\cdot{\bm \sigma})\hat{\bm \Omega}_{{\bm k}'}-\hat{\bm \Omega}_{\bm k}(\hat{\bm \Omega}_{{\bm k}'}\cdot{\bm \sigma})\Big]\Big[\delta(\epsilon_+-\epsilon'_-)+\delta(\epsilon_--\epsilon'_+)\Big]\\[3ex]
  &\dps +\frac{\pi n_i}{4\hbar}\int\frac{d^2k'}{(2\pi)^2}|\overline{U}_{{\bm k}{\bm k}'}|^2({\bm S}_{\bm k}-{\bm S}_{{\bm k}'})
  \cdot\Big[(\hat{\bm \Omega}_{\bm k}-\hat{\bm \Omega}_{{\bm k}'})
\Big]\Big[\delta(\epsilon_+-\epsilon'_-)-\delta(\epsilon_--\epsilon'_+)\Big].
   \ea
  \end{equation} 
 We now separate these terms according to the contributions from  intra-band and inter-band scatterings
   \begin{equation}
   \ba
  \hat{J}(n)  &\dps =\frac{\pi n_i}{2\hbar}\int\frac{d^2k'}{(2\pi)^2}|\overline{U}_{{\bm k}{\bm k}'}|^2(n_{\bm k}-n_{{\bm k}'})
  (1+\cos3\gamma)\Big[\delta(\epsilon_+-\epsilon'_+)+\delta(\epsilon_--\epsilon'_-)\Big]\\[3ex]
    &\dps +\frac{\pi n_i}{2\hbar}\int\frac{d^2k'}{(2\pi)^2}|\overline{U}_{{\bm k}{\bm k}'}|^2(n_{\bm k}-n_{{\bm k}'})
 (1-\cos3\gamma)\Big[\delta(\epsilon_+-\epsilon'_-)+\delta(\epsilon_--\epsilon'_+)\Big],
   \ea
  \end{equation}
   \begin{equation}
   \ba 
  \hat{J}(S)
   &\dps =\frac{\pi n_i}{4\hbar}\int\frac{d^2k'}{(2\pi)^2}|\overline{U}_{{\bm k}{\bm k}'}|^2
  \Big[(s_{{\bm k}\parallel}-s_{{\bm k}'\parallel})(1+\cos3\gamma)\sigma_{{\bm k}\parallel}+(s_{{\bm k}\parallel}-s_{{\bm k}'\parallel})\sin3\gamma\sigma_{{\bm k}\perp}\\[3ex]
  &\dps +(s_{{\bm k}\perp}+s_{{\bm k}'\perp})\sigma_{{\bm k}\parallel}\sin3\gamma +(s_{{\bm k}\perp}+s_{{\bm k}'\perp})(1-\cos3\gamma)\sigma_{{\bm k}\perp}\Big]\Big[\delta(\epsilon_+-\epsilon'_+)+\delta(\epsilon_--\epsilon'_-)\Big] \\[3ex]
  &\dps +\frac{\pi n_i}{4\hbar}\int\frac{d^2k'}{(2\pi)^2}|\overline{U}_{{\bm k}{\bm k}'}|^2
 \Big[(s_{{\bm k}\parallel}+s_{{\bm k}'\parallel})(1-\cos3\gamma)\sigma_{{\bm k}\parallel}-(s_{{\bm k}\parallel}+s_{{\bm k}'\parallel})\sin3\gamma\sigma_{{\bm k}\perp}\\[3ex]
 &\dps -(s_{{\bm k}\perp}-s_{{\bm k}'\perp})\sigma_{{\bm k}\parallel}\sin3\gamma+(s_{{\bm k}\perp}-s_{{\bm k}'\perp})(1-\cos3\gamma)\sigma_{{\bm k}\perp}\Big]\Big[\delta(\epsilon_+-\epsilon'_-)+\delta(\epsilon_--\epsilon'_+)\Big],
 \ea
  \end{equation}
  and
   \begin{equation}
   \ba 
\hat{J}_{S\rightarrow n}(S) &\dps =\frac{\pi n_i}{4\hbar}\int\frac{d^2k'}{(2\pi)^2}|\overline{U}_{{\bm k}{\bm k}'}|^2
  \Big[(s_{{\bm k}\parallel}-s_{{\bm k}'\parallel})(1+\cos3\gamma)+(s_{{\bm k}\perp}+s_{{\bm k}'\perp})\sin3\gamma\Big]  
  \Big[\delta(\epsilon'_+-\epsilon_+)-\delta(\epsilon'_--\epsilon_-)\Big] \\[3ex]
  &\dps =\frac{\pi n_i}{4\hbar}\int\frac{d^2k'}{(2\pi)^2}|\overline{U}_{{\bm k}{\bm k}'}|^2
  \Big[(s_{{\bm k}\parallel}+s_{{\bm k}'\parallel})(1-\cos3\gamma)-(s_{{\bm k}\perp}-s_{{\bm k}'\perp})\sin3\gamma\Big]  
  \Big[\delta(\epsilon_+-\epsilon'_-)-\delta(\epsilon_--\epsilon'_+)\Big],
   \ea
  \end{equation}
   \begin{equation}
   \ba 
   \hat{J}_{n\rightarrow S}(n)  &\dps =\frac{\pi n_i}{2\hbar}\int\frac{d^2k'}{(2\pi)^2}|\overline{U}_{{\bm k}{\bm k}'}|^2(n_{\bm k}-n_{{\bm k}'})\Big[\sigma_{{\bm k}\parallel}(1+\cos3\gamma)+\sigma_{{\bm k} \perp}\sin3\gamma\Big]\Big[\delta(\epsilon'_+-\epsilon_+)-\delta(\epsilon'_--\epsilon_-)\Big]\\[3ex]
   &\dps +\frac{\pi n_i}{2\hbar}\int\frac{d^2k'}{(2\pi)^2}|\overline{U}_{{\bm k}{\bm k}'}|^2(n_{\bm k}-n_{{\bm k}'})\Big[\sigma_{{\bm k}\parallel}(1-\cos3\gamma)-\sigma_{{\bm k} \perp}\sin3\gamma\Big]\Big[\delta(\epsilon'_--\epsilon_+)-\delta(\epsilon'_+-\epsilon_-)\Big].
    \ea
  \end{equation}
We next decompose the kinetic equation as follows:
\begin{equation}\label{3K}
\ba
&\dps \frac{d n_{\bm k}}{dt}+\hat{J}_{n\rightarrow n} (n_{\bm k}) =\mathcal{D}_{{\bm k}n},\\[3ex]
&\dps \frac{d S_{{\bm k}\parallel}}{dt}+P_{\parallel}\hat{J}_{S \rightarrow S} (S_{{\bm k}\parallel})
=\mathcal{D}_{{\bm k}\parallel},\\[3ex]
&\dps \frac{d S_{{\bm k}\perp}}{dt}+\frac{i}{\hbar}\big[H_{0{\bm k}},S_{{\bm k}\perp}\big]=\mathcal{D}_{{\bm k}\perp}.
\ea
\end{equation}
Note that the projection operator $P_{\parallel}$ above acts on a matrix $\mathcal{M}$ as $\text{Tr}(\mathcal{M}\sigma_{{\bm k}\parallel})$, where Tr refers to the matrix (spin) trace.
   
\section{Solution for the longitudinal conductivity}\label{E}
Here we derive the longitudinal conductivity at zero magnetic field. Expanding the $\delta$ functions in Sec. \ref{J} up to $\propto\alpha^{2}$, we get the following
\begin{equation}\label{delta}
 \ba
&\dps  \delta(\epsilon_+-\epsilon'_+)\approx \delta(\epsilon_{0}-\epsilon'_{0})+\alpha(k^3-k'^3)\pd{}{\epsilon_0}\delta(\epsilon_{0}-\epsilon'_{0})+\frac{\alpha^2(k^3-k'^3)^2}{2}\frac{\partial^{2} \delta(\epsilon_0-\epsilon'_0)}{\partial \epsilon^{2}_0} \\[3ex]
&\dps  \delta(\epsilon_--\epsilon'_-)\approx \delta(\epsilon_0-\epsilon'_0)-\alpha(k^3-k'^3)\pd{}{\epsilon_0}\delta(\epsilon_0-\epsilon'_0)+\frac{\alpha^2(k^3-k'^3)^2}{2}\frac{\partial^{2} \delta(\epsilon_0-\epsilon'_0)}{\partial \epsilon^{2}_0}\\[3ex]
&\dps  \delta(\epsilon_+-\epsilon'_-)\approx \delta(\epsilon_{0}-\epsilon'_{0})+\alpha(k^3+k'^3)\pd{}{\epsilon_0}\delta(\epsilon_{0}-\epsilon'_{0})+\frac{\alpha^2(k^3+k'^3)^2}{2}\frac{\partial^{2} \delta(\epsilon_0-\epsilon'_0)}{\partial \epsilon^{2}_0} \\[3ex]
&\dps  \delta(\epsilon_--\epsilon'_+)\approx \delta(\epsilon_0-\epsilon'_0)-\alpha(k^3+k'^3)\pd{}{\epsilon_0}\delta(\epsilon_0-\epsilon'_0)+\frac{\alpha^2(k^3+k'^3)^2}{2}\frac{\partial^{2} \delta(\epsilon_0-\epsilon'_0)}{\partial \epsilon^{2}_0}.
\ea
\end{equation}  
We now insert Eq.~(\ref{delta}) into the electric driving term $\mathcal{D}_{E,{\bm k}}$ and scattering term $\hat{J}(f_{\bm k})$.  With 
$\rho_{0{\bm k}}=\frac{f_{0+}+f_{0-}}{2}\openone+\frac{f_{0+}-f_{0-}}{2}\sigma_{{\bm k}\parallel}$ and $f_{0+}$, $f_{0-}$ equilibrium Fermi distribution function, the driving term $\mathcal{D}_{E,{\bm k}}$ becomes,
\begin{equation}
\ba
&\dps  \mathcal{D}_{E,{\bm k}n}=-\frac{e{\bm E}\cdot\hat{\bm k}}{2\hbar}(\pd{f_{0+}}{k}+\pd{f_{0-}}{k})
\approx\frac{e{\bm E}\cdot\hat{\bm k}}{2\hbar}\Big[2\frac{\hbar^2 k}{m^*}\delta(\epsilon_0-\epsilon_{\text{F}})+6\alpha^2k^5\pd{\delta(\epsilon_0-\epsilon_{\text{F}})}{\epsilon_0}\Big],\\[3ex]
&\dps  \mathcal{D}_{E,{\bm k}\parallel}=-\frac{e{\bm E}\cdot\hat{\bm k}}{2\hbar}(\pd{f_{0+}}{k}-\pd{f_{0-}}{k})\sigma_{{\bm k}\parallel}\approx \frac{e{\bm E}\cdot\hat{\bm k}}{2\hbar}\Big[6\alpha k^2\delta(\epsilon_0-\epsilon_{\text{F}})+2\frac{\hbar^2 k}{m^*}\alpha k^3\pd{\delta(\epsilon_0-\epsilon_{\text{F}})}{\epsilon_0})\Big].
\ea
\end{equation}
Solving Eqs.~(\ref{3K}), we obtain the density matrices 
\begin{subequations}
\begin{equation}
\ba
 &\dps  n^{(0)}_{E{\bm k}}=\tau_p\frac{e{\bm E}\cdot \hat{\bm k}}{\hbar}\Big[\frac{\hbar^2k}{m^*}\delta(\epsilon_0-\epsilon_{\text{F}})\Big],
\ea
\end{equation}
\begin{equation}
\ba
S^{(1)}_{E{\bm k}\parallel}
&\dps  =\tau_s \alpha \frac{e{\bm E}\cdot\hat{\bm k}}{\hbar}\Big[\frac{\hbar^2k^4}{m^*}\pd{\delta(\epsilon_0-\epsilon_{\text{F}})}{\epsilon_0}+3 k^2\delta(\epsilon_0-\epsilon_{\text{F}})\Big]\sigma_{{\bm k}\parallel}=s^{(1)}_{E{\bm k}\parallel}\sigma_{{\bm k}\parallel},
\ea
\end{equation} 
\begin{equation}
\ba
 n^{(2)}_{E{\bm k}} &\dps = \tau_p \alpha^2\Big\{\frac{e{\bm E}\cdot\hat{\bm k}}{\hbar}\Big[3k^5\pd{\delta(\epsilon_0-\epsilon_{\text{F}})}{\epsilon_0}\Big]-\frac{3 k m^{*2}n_i}{4\alpha \pi\hbar^5 }s^{(1)}_{E{\bm k}\parallel} \zeta(\gamma) -n^{(0)}_{E{\bm k}}\frac{6n_im^{*3}}{\pi\hbar^7}k^2\xi(\gamma)\Big\}.
 \ea
\end{equation}
\end{subequations}
where $\epsilon_\text{F}=\frac{\hbar^2k^2_{\text{F}}}{2m^*}$, $\tau_p=\frac{2\pi\hbar^3}{m^*n_i\xi(\gamma)}$, $\tau_s=\frac{4\pi\hbar^3}{m^*n_i\beta(\gamma)}$, and 
\begin{equation}
\ba  
&\dps \zeta(\gamma)\!=\!\int d\gamma|\overline{U}_{{\bm k}{\bm k}'}|^2(\cos\gamma-\cos3\gamma),\quad \xi(\gamma)\!=\!\int d\gamma |\overline{U}_{{\bm k}{\bm k}'}|^2(1-\cos\gamma),\quad \beta(\gamma)\!=\!\int d\gamma |\overline{U}_{{\bm k}{\bm k}'}|^2(1-\cos\gamma\cos3\gamma).
\ea
\end{equation} 

In the low temperature limit, the Thomas-Fermi wave-vector of a two-dimensional hole gas without spin-orbit coupling is $k_{\text{TF}}=\frac{2}{a_{\text{B}}}$, with $a_{\text{B}}=\frac{\hbar^2\epsilon_r}{m^*e^2}$.  The screened Coulomb potential between plane waves is given by
\begin{equation}\label{screened-C}
|\overline{U}_{{\bm k}{\bm k}'}|^2=\frac{Z^2e^4}{4\epsilon^2_0\epsilon^2_r}\left(\frac{1}{|{\bm k}-{\bm k}'|+k_{\text{TF}}}\right)^2.
\end{equation}
With Eq.~(\ref{screened-C}), we obtain $\frac{ \zeta(\gamma)}{ \xi(\gamma)}\approx 2$ and ${\frac{\xi(\gamma)}{\beta(\gamma)}}=\frac{1}{3}$. Using the velocity operator 
\begin{equation}\label{vx}
\ba
&\dps \hat{v}_x=\frac{\hbar k_x}{m^*}+\frac{\alpha}{\hbar}3k^2[-\sin2\theta\sigma_x+\cos2\theta\sigma_y],\quad \hat{v}_y=\frac{\hbar k_y}{m^*}+\frac{\alpha}{\hbar}3k^2[-\sin2\theta\sigma_y-\cos2\theta\sigma_x],
\ea
\end{equation}
the longitudinal current is $j_x=e\text{Tr}\big[\hat{v}_x\rho_{E{\bm k}}\big]$, where $\rho_{E{\bm k}}=(n^{(0)}_{E{\bm k}}+n^{(2)}_{E{\bm k}})\openone+ S^{(1)}_{E{\bm k}\parallel}$. Therefore, the longitudinal conductivity with Rashba spin orbit coupling up to second order in $\alpha$ is
\begin{equation}
\sigma_{xx}=\frac{e^2\tau_p}{2\pi m^*}k^2_{\text{F}}\Big[1-\frac{15}{2}\left(\frac{\alpha k^3_{\text{F}}}{\epsilon_{kin}}\right)^2\Big],
\end{equation} 
where $\epsilon_{kin}=\frac{\hbar^2k^2_{\text{F}}}{2m^*}$.

\section{Solution for the Hall coefficient}\label{H}
Now we consider the case of $B_z>0$. Firstly, we find that the Zeeman terms have no contribution to the Hall coefficient. With Eqs.~(\ref{vx}), the Lorentz driving term $\mathcal{D}_{L,{\bm k}}$ becomes 
\begin{equation}\label{L-D}
\mathcal{D}_{L,{\bm k}}=\frac{1}{2}\frac{e}{\hbar }\Big\{\hat{\bm v}\times {\bm B}, \pd{\rho_{E{\bm k}}}{{\bm k}}\Big\}=\frac{1}{2}\frac{eB_z}{\hbar }\Big\{\big\{\hat{v}_y, \pd{\rho_{E{\bm k}}}{k_x}\big\}-\big\{\hat{v}_x,\pd{\rho_{E{\bm k}}}{k_y}\big\}\Big\}.
\end{equation}
We separate $\mathcal{D}_{L,{\bm k}}$ into the scalar and spin parts with $\mathcal{D}_{L,{\bm k}}=\mathcal{D}_{L,n}+\mathcal{D}_{L,S}$, {and, switching from the rectangular coordinates to polar coordinates with $\pd{\mathcal{D}}{k_x}=\pd{\mathcal{D}}{k}\cos\theta-\pd{\mathcal{D}}{\theta}\frac{\sin\theta}{k};\pd{\mathcal{D}}{k_y}=\pd{\mathcal{D}}{k}\sin\theta+\pd{\mathcal{D}}{\theta}\frac{\cos\theta}{k}$, we obtain 
 \begin{equation}
\ba
&\dps \mathcal{D}_{L,n}=-\frac{eB_z}{m^*}\big[n^{(0)}_k+n^{(2)}_k\big](-\sin\theta)+\frac{eB_z}{\hbar}\frac{3\alpha k}{\hbar}s^{(1)}_{k,\parallel}(-\sin\theta),\\[3ex]
&\dps \mathcal{D}_{L,S_{\parallel}}=-\Big\{\frac{eB_z}{m^*}s^{(1)}_{k,\parallel}(-\sin\theta)+\frac{eB_z}{\hbar}\frac{3\alpha k}{\hbar}\big[n^{(0)}_k+n^{(2)}_k\big](-\sin\theta)\Big\}\sigma_{{\bm k}\parallel},\\[3ex]
&\dps \mathcal{D}_{L,S_{\perp}}=\cos\theta\frac{eB_z}{\hbar}\frac{3\alpha k^2}{\hbar}\pd{\big[n^{(0)}_k+n^{(2)}_k\big]}{k}\sigma_{{\bm k}\perp},
\ea
\end{equation}
with $n^{(0)}_{E{\bm k}}=n^{(0)}_k \cos\theta $, $n^{(2)}_{E{\bm k}}= n^{(2)}_k \cos\theta$ and $s^{(1)}_{E{\bm k},\parallel}= s^{(1)}_{k,\parallel} \cos\theta$. Solving Eqs.~(\ref{3K}), we obtain the following density matrices in presence both electric and magnetic fields
\begin{equation}
\ba
&\dps {n}_{B_z,{\bm k}}=-\sin\theta\tau_peB_z\Big\{\frac{n^{(0)}_k+n^{(2)}_k}{m^*}+\frac{3\alpha k}{\hbar^2}s^{(1)}_{k,\parallel}\Big\},\\[3ex]
&\dps {S}_{B_z,{\bm k}\parallel}=-\sin\theta\tau_seB_z\Big\{\frac{s^{(1)}_{k,\parallel}}{m^*}+\frac{3\alpha k}{\hbar^2}\big[n^{(0)}_k+n^{(2)}_k\big]\Big\}\sigma_{{\bm k}\parallel},\\[3ex]
&\dps S_{B_z,{\bm k}\perp}=\cos\theta\frac{3eB_z}{2\hbar k}\pd{\big[n^{(0)}_k+n^{(2)}_k\big]}{k}\sigma_z.
\ea
\end{equation}

The Hall current is $ j_y=e\text{Tr}\big[\hat{v}_y\rho^{EB}_{\bm k}\big]$, where $\rho^{EB_z}_{\bm k}={n}_{B_z,{\bm k}}\openone+{S}_{B_z,{\bm k}\parallel}+S_{B_z,{\bm k}\perp}$. The Hall coefficient, up to the second order in $\alpha$, is thus given as
\begin{equation}
R_H=\frac{\sigma_{xy}}{B_z(\sigma^2_{xx} + \sigma^2_{xy})}\approx\frac{1}{pe}\Big[1+8\left(\frac{\alpha k^3_{\text{F}}}{\epsilon_{kin}}\right)^2\Big],
\end{equation}
where $\omega_c=\frac{eB_z}{m^*}$.